\documentclass[12pt]{article}
\usepackage{geometry}                % See geometry.pdf to learn the layout options. There are lots.
\usepackage{graphicx}
\usepackage{bm}
\usepackage{amssymb,amsmath,amsthm}
\usepackage{mathrsfs} %script fonts
\def\TODAY{9 January 2010; 30 April 2010}
\title{\bf Acoustic geometry for general relativistic barotropic irrotational fluid flow}
\author{{\bf Matt Visser}\\
School of Mathematics, Statistics, and Operations Research, \\
Victoria University of Wellington, \\
Wellington, New Zealand\\[10pt]
and\\[10pt]
{\bf Carmen Molina--Par\'is}\\
Department of Applied Mathematics, \\
University of Leeds, \\
Leeds LS2 9JT, UK 
}
\date{\TODAY; \LaTeX-ed \today}                                           
\begin{document}
%------------------------------------------------------------------------------------------------------------------------------------------
\maketitle
%------------------------------------------------------------------------------------------------------------------------------------------
% very standard definitions
%------------------------------------------------------------------------------------------------------------------------------------------
\def\d{{\mathrm{d}}}
\newcommand{\scri}{\mathscr{I}}
\newcommand{\sun}{\ensuremath{\odot}}
%------------------------------------------------------------------------------------------------------------------------------------------
%------------------------------------------------------------------------------------------------------------------------------------------
%------------------------------------------------------------------------------------------------------------------------------------------
%------------------------------------------------------------------------------------------------------------------------------------------

\begin{abstract}
``Acoustic spacetimes'', in which techniques of
differential geometry are used to investigate sound propagation in
moving fluids, have attracted considerable attention over the last few
decades. Most of the models currently considered in the literature are based on
non-relativistic barotropic irrotational fluids, defined in a flat
Newtonian background. The extension, first to special relativistic barotropic
fluid flow, and then to general relativistic barotropic fluid flow in an
arbitrary background, is less straightforward than it might at first
appear. In this article we provide a pedagogical
and simple derivation of the general relativistic ``acoustic
spacetime'' in an arbitrary $(d+1)$ dimensional curved-space
background.


\bigskip
\noindent
%File: {\sf \jobname .tex}
\end{abstract}

%------------------------------------------------------------------------------------------------------------------------------------------
\clearpage
%------------------------------------------------------------------------------------------------------------------------------------------
\hrule
%------------------------------------------------------------------------------------------------------------------------------------------
\tableofcontents
%------------------------------------------------------------------------------------------------------------------------------------------

\bigskip
\hrule

\section{Introduction}
%------------------------------------------------------------------------------------------------------------------------------------------
\def\G{{\mathcal{G}}}
%------------------------------------------------------------------------------------------------------------------------------------------

In this article we shall present a simple
and pedagogical
derivation of the general relativistic version of the ``acoustic
metric'' defined on an arbitrary curved $(d+1)$ dimensional background
spacetime.  While there are related observations and more limited
derivations extant in the literature, we feel that the current
analysis has some definite advantages.  For instance, the early 1980
analysis due to Moncrief is restricted to perturbations of a
spherically symmetric fluid flow on a Schwarzschild
background~\cite{Moncrief}, and in the more recent 1999 derivation due
to Bilic~\cite{Bilic} it can be somewhat difficult to discern what is
truly fundamental input from what is derived output. Since scientific
interest in this field is both significant and ongoing~\cite{LRR, Stefano-pvt,  Silke-pvt}, we feel it useful to carefully lay out the
minimal set of assumptions and logic flow behind the derivation. 

\bigskip
\noindent
Our
strategy is as follows:
\begin{itemize}
\item 
We shall first motivate the result (up to a conformal factor) by considering the acoustic  version of the Gordon metric, which was  introduced for geometrical optics in 1923~\cite{Gordon}. 
\item
We shall carefully specify what we mean by barotropic flow, the assumption that energy density is a function of pressure only. (Specifically, barotropic flow is more general than either isothermal or isentropic flow.) 

\item 
We shall then carefully specify what is meant by ``irrotational
flow'' in a general relativistic context, introducing the appropriate notion of velocity potential.
\item 
From the relativistic Euler equation, using only the irrotational
condition and the barotropic condition, we will derive the
relativistic Bernoulli equation.
\item 
From the relativistic energy equation, using only the barotropic
condition, (that is, without using the irrotational condition), we
shall derive a flux conservation law (continuity equation).
\item 
We shall delay the introduction of thermodynamic arguments as long as
practicable. (We would argue that thermodynamics is in fact a side issue not central to the derivation.) 
\item
As usual, the acoustic metric follows from combining the linearized
Bernoulli equation and linearized equation of state with the
linearized continuity equation.
\item 
We shall carefully explain the subtleties involved in taking the non-relativistic limit.
\item
We shall finish with some discussion, and relegate several
thermodynamic observations to the appendices.
\end{itemize}

\bigskip
\noindent
The key result that we shall be aiming for is this: the
(contravariant) acoustic metric governing acoustic perturbations of an
irrotational barotropic fluid flow in $(d+1)$ dimensions is
 \begin{equation}
\G^{ab} = \left( {n_0^2\; c_s^{-1} \over\varrho_0+p_0} \right)^{-2/(d-1)}  \\
 \left\{   -{c^2\over c_s^2} \; V_0^a \, V_0^b  +h^{ab} \right\}.
 \end{equation}
In counterpoint, the (covariant) acoustic metric in $(d+1)$ dimensions is
 \begin{equation}
\G_{ab} =  \;\;\left( {n_0^2\; c_s^{-1}\over\varrho_0+p_0} \right)^{2/(d-1)}  \\
 \left\{   -{ c_s^2\over c^2}  \; [V_0]_a  \, [V_0]_b  +h_{ab} \right\}.
 \end{equation}

%\newpage
\noindent
Here:
\begin{itemize}
\item $c_s$ is the speed of sound, defined as usual via $c_s^2 = c^2 \; \partial p/\partial\varrho$.
\item $c$ is the speed of light, used for instance in defining $x^0 = c\,t$.
\item $V_0$ is the dimensionless 4-velocity of the background fluid flow.
\item $h_{ab} = g_{ab} + [V_0]_a [V_0]_b $ is the dimensionless orthogonal
  projection of the physical spacetime metric $g_{ab}$ onto the
  3-space perpendicular to the 4-velocity of the fluid.
\item the indices on the background fluid flow $[V_0]^a$ are lowered and raised using the physical spacetime metric $g_{ab}$ and its inverse $g^{ab}$. 
\item in contrast $\G_{ab}$ and $\G^{ab}$ are defined to be matrix inverses of each other; these indices are \emph{not} to be raised and lowered with the physical spacetime metric.
\item $n_0$ is the background number density of fluid particles.
\item $\varrho_0$ is the background energy density of the fluid.
\item $p_0$ is the background pressure of the fluid.
\end{itemize}

\bigskip
\noindent
If one appeals to relativistic thermodynamics, by introducing the specific enthalpy $w = (\varrho+p)/n$, then, as discussed in the appendix, the acoustic metric can be somewhat simplified:
 \begin{equation}
\G^{ab} = \left( {n_0\over w_0 \;  c_s} \right)^{-2/(d-1)}  \\
 \left\{   -{c^2\over c_s^2} \; V_0^a \, V_0^b  +h^{ab} \right\};
 \end{equation}
 \begin{equation}
\G_{ab} =  \;\;\left( {n_0\over w_0  \; c_s} \right)^{2/(d-1)}  \\
 \left\{   -{ c_s^2\over c^2}  \; [V_0]_a  \, [V_0]_b  +h_{ab} \right\}.
 \end{equation}

\bigskip
\noindent
These key results are easy to \emph{motivate} (but not \emph{derive}) in
the limit of ``ray acoustics'', also known as ``geometric acoustics'',
where we can safely ignore the wave properties of sound. In this limit
we are interested only in the ``sound cones''. Let us pick a point in
spacetime where the background fluid 4-velocity is $V_0^a$. Now adopt
Gaussian normal coordinates, \emph{and} go to the local rest frame of the
fluid. Then taking $x^0 = ct$ we have
\begin{equation}
[V_0]^a \to (1; 0,0,0),
\end{equation}
and
\begin{equation}
g_{ab} \to \left[\begin{array}{cccc}-1&0&0&0\\0&1&0&0\\0&0&1&0\\0&0&0&1\end{array} \right];
\qquad
h_{ab} = g_{ab} + [V_0]_a\, [V_0]_b \to \left[\begin{array}{cccc}0&0&0&0\\0&1&0&0\\0&0&1&0\\0&0&0&1\end{array} \right].
\end{equation}
In the rest frame of the fluid the sound cones are (locally) given by
\begin{equation}
- c_s^2 \; \d t^2 + ||\d \vec x||^2 = 0,
\end{equation}
which we can rewrite as
\begin{equation}
- {c_s^2\over c^2}  \; (c \,\d t)^2 + ||\d \vec x||^2 =  - {c_s^2\over c^2}  \; (\d x^0)^2 + ||\d \vec x||^2=   0,
\end{equation}
implying in these special coordinates the existence of an acoustic
metric
\begin{equation}
\G_{ab} \propto  \left[\begin{array}{cccc}-c_s^2/c^2&0&0&0\\0&1&0&0\\0&0&1&0\\0&0&0&1\end{array} \right].
\end{equation}
That is, transforming back to arbitrary coordinates:
\begin{equation}
\G_{ab} \propto -{c_s^2\over c^2} \;  [V_0]_a\, [V_0]_b + h_{ab}.
\end{equation}
We now rewrite this as
\begin{equation}
\G_{ab} \propto -{c_s^2\over c^2}  [V_0]_a\, [V_0]_b + \left\{ g_{ab} + [V_0]_a\, [V_0]_b\right\}  \propto g_{ab} + \left\{1-{c_s^2\over c^2}\right\} [V_0]_a \,[V_0]_b.
\end{equation}
Note that this is essentially a generalization of the derivation (dating
back to 1923) of the so-called ``Gordon metric''~\cite{Gordon} used in
ray optics to describe the ``optical metric'' appropriate to a
(possibly) relativistic fluid with position-dependent refractive index
$n(t,\vec x)$. In Gaussian normal coordinates comoving with the fluid
the Gordon metric is given by
\begin{equation}
\G_{ab} \propto  \left[\begin{array}{cccc}-{1/ n^2} &0&0&0\\0&1&0&0\\0&0&1&0\\0&0&0&1\end{array} \right].
\end{equation}
That is, transforming back to arbitrary coordinates:
\begin{equation}
\G_{ab} \propto -{1\over n^2}  [V_0]_a\, [V_0]_b + \left\{ g_{ab} + [V_0]_a\, [V_0]_b\right\}  
\propto g_{ab} + \left\{1-{1\over n^2}\right\} [V_0]_a\, [V_0]_b.
\end{equation}
Note that in either the ray acoustics or ray optics limits, because one
only has the sound cones or light cones to work with, one can neither
derive nor is it even meaningful to specify the overall conformal
factor~\cite{LRR, unexpected, ergo}. (See also reference~\cite{Carter} for an equivalent Godon-like metric based on relativistic fluid dynamics but without calculation of the overall conformal factor.) The calculation presented below
is designed to go beyond the ray acoustics limit, to obtain a
relativistic wave equation suitable for describing physical acoustics
--- all the ``fuss'' is simply over how to determine the overall
conformal factor (and to verify that one truly does obtain a
d'Alembertian equation using the conformally fixed acoustic metric).

%------------------------------------------------------------------------------------------------------------------------------------------
\section{Basic general relativistic fluid mechanics}
%------------------------------------------------------------------------------------------------------------------------------------------

In this Section we consider fully relativistic barotropic inviscid
irrotational flow on an arbitrary general relativistic background, and
derive the relevant wave equation for linearized fluctuations. We
already know exactly what happens in the non-relativistic
case~\cite{LRR, Unruh, unexpected, ergo}. We 
go straight to curved spacetime relativistic fluid mechanics, where
the geometry is described by a metric tensor $g_{ab}(x)$ of signature
$-+++$, [or more generally $-(+)^d$], and the fluid is described by
the energy density $\varrho$, pressure $p$, and 4-velocity $V^a$ where
\begin{equation}
g_{ab} V^a V^b = -1.
\end{equation}
The two relevant fluid dynamical equations are extremely well-known (see, for instance, Hawking and Ellis~\cite{Hell}).
\begin{description}
\item[Relativistic energy equation:] 
\begin{equation}
\nabla_a ( \varrho V^a) + p (\nabla_a V^a) = 0.
\end{equation}
\item[Relativistic Euler equation:] 
\begin{equation}
(\varrho+p) V^b \nabla_b V^a = - (g^{ab}+V^a V^b)\nabla_b p.
\end{equation}
\end{description}
These equations can be combined into the single statement that the stress-energy
tensor is covariantly conserved (see, for instance, Hawking and Ellis~\cite{Hell})
\begin{equation}
\nabla_a \left[ (\varrho+p) V^a V^b + p g^{ab} \right] = 0.
\end{equation}
Note the subtle (perhaps not so subtle) differences from the
non-relativistic case: $\varrho$ is now energy density, \emph{not mass
  density $\rho$}, and the continuity equation looks different --- at
least at this stage of the calculation it has been replaced by an
``energy conservation'' equation, and the Euler equation now contains
the combination $(\varrho+p)$. The 4-acceleration of the fluid is
\begin{equation}
A^a = V^b \nabla_b V^a,
\qquad \text{such that}  \qquad   A^a \; V_a = 0.
\end{equation}

%------------------------------------------------------------------------------------------------------------------------------------------
\subsection{Defining relativistic barotropic  flow}
%------------------------------------------------------------------------------------------------------------------------------------------

The fluid is barotropic if the energy density is a function only of the pressure --- that is, there is some function such that $\varrho = \varrho(p)$.  Appealing to the inverse function theorem we will have $p = p(\varrho)$ at least piecewise on suitable open sets in energy density.  This barotropic condition is much more general situation than assuming isothermal or isentropic flow. Specifically, for a one-component fluid:
\begin{itemize}
\item 
Any zero-temperature fluid is automatically barotropic. 
\item
Any non-zero temperature but isothermal fluid is automatically barotropic. 
\item
Any zero-entropy fluid (superfluid) is automatically barotropic.
\item 
Any isentropic fluid is automatically barotropic. 
\end{itemize}
However, merely assuming the barotropic condition does not imply any of the isentropic, zero-entropy, isothermal, or zero-temperature conditions.

%------------------------------------------------------------------------------------------------------------------------------------------
\subsection{Defining relativistic irrotational flow}
%------------------------------------------------------------------------------------------------------------------------------------------

What do we mean by irrotational flow in a relativistic setting?
Construct the 1-form
\begin{equation}
v = V_a \; \d x^a,
\end{equation}
and consider the 2-form $\omega_2$ and 3-form $t_3$:
\begin{equation}
\omega_2 = \d v; \qquad t_3 = v \wedge \omega_2.
\end{equation}
In a relativistic context, setting the 2-form $\omega_2=0$ is too
strong a condition, setting the 3-form $t_3=0$ is just right. (See, for instance, Hawking and Ellis~\cite{Hell}.) Indeed,
adopting Gaussian normal coordinates and going to the local rest frame of
the fluid, where $V^a\to(1;0,0,0)$, setting $t_3=0$ implies
$\partial_{[i} v_{j]} = 0$, so the \emph{spatial} components of the
flow velocity are locally irrotational (in the sense of being curl-free).  But $t_3=0$ implies (via the
Frobenius theorem) that locally there exist functions $\alpha$ and $\Theta$ such that
 \begin{equation}
 v = \alpha \; \d\Theta; \qquad V^a = \alpha \; g^{ab} \;\nabla_b \Theta.
 \end{equation}
 But then, from the normalization condition for the 4-velocity
 \begin{equation}
 V^a   =  {g^{ab} \; \nabla_b \Theta\over\sqrt{- g^{ab} \;\nabla_a \Theta\; \nabla_b \Theta}}.
 \end{equation}
We shall find it extremely useful to define
\begin{equation}
||\nabla \Theta||^2 = - g^{ab} \;\nabla_a \Theta\; \nabla_b \Theta
\end{equation}
so that
 \begin{equation}
 V^a  =
 {g^{ab} \; \nabla_b \Theta\over ||\nabla \Theta||}.
 \end{equation}
This is the \emph{relativistic} condition for irrotational flow. The function  $\Theta$ can now be interpreted as the general relativistic version of the velocity potential. Note
that for any smooth function $F(\cdot)$ we can replace $\Theta
\leftrightarrow F(\Theta)$ \emph{without affecting the 4-velocity}
$V^a$ --- both numerator and denominator above pick up factors of
$F'(\Theta)$ which then cancel.  This freedom in choosing the scalar
potential, $\Theta$, will be very useful when analyzing the Euler
equation, and will allow us to obtain a relativistic Bernoulli
equation.

For the discussion below define the projection operator
\begin{equation}
h^{ab} = g^{ab} + V^a \, V^b.
\end{equation}
Using this projection operator, and the 4-orthogonality of 4-velocity
with 4-acceleration, it is easy (for relativistic irrotational flow) to
calculate
\begin{equation}
\nabla_b V_a  = (\delta_a{}^c + V_a V^c )  {\nabla_c \nabla_b \Theta
\over||\nabla \Theta||}.
\end{equation}
Therefore, for relativistic irrotational flow, the 4-acceleration
reduces to
\begin{equation}
A_a = V^b \nabla_b V_a =
 - {1\over2} (\delta_a{}^c + V_a V^c ) 
 {\nabla_c (- g_{de} \nabla_d \Theta \nabla_e\Theta )
\over {- g^{ab} \;\nabla_a \Theta\; \nabla_b \Theta}},
\end{equation}
so that
\begin{equation}
A^a = -(g^{ab} + V^a V^b ) 
 {\nabla_b \left(\log||\nabla \Theta|| \right)
} .
\end{equation}
Note that automatically $A^a V_a=0$, as required.  Furthermore, note
that the 4-acceleration is the \emph{projection} of the gradient of a
scalar, and that $\Theta$ can be chosen to have  the dimensions of a distance so that $\nabla\Theta$ is dimensionless.
 
%------------------------------------------------------------------------------------------------------------------------------------------
\subsection{From Euler equation to Bernoulli equation}
%------------------------------------------------------------------------------------------------------------------------------------------

The Euler equation for relativistic irrotational flow now reads
 \begin{equation}
  (g^{ab} + V^a V^b ) \;
 \nabla_b \left( \log||\nabla \Theta|| \right)
=
  (g^{ab}+V^a V^b)\; {\nabla_b p\over\varrho+p}.
 \end{equation}
We now make use of the barotropic condition $\varrho=\varrho(p)$ to obtain
\begin{equation}
{\nabla_b p\over\varrho+p} = \nabla_b \int_0^p {\d p\over\varrho(p)+p},
\end{equation}
so the Euler equation becomes
\begin{equation}
 (g^{ab} + V^a V^b ) 
 \nabla_b \left( - \log||\nabla \Theta||
 + \int_0^p {\d p\over\varrho(p)+p}\right) = 0.
\end{equation}
Note that for any arbitrary function $f(\Theta)$ we have
\begin{equation}
  (g^{ab} + V^a V^b )  \nabla_b f(\Theta) = 0,
\end{equation}
(since the projection operator kills the gradient).
Therefore (using both the irrotational and barotropic conditions) we can integrate the Euler equation to yield:
\begin{equation}
-\log||\nabla \Theta||
 + \int_0^p {\d p\over\varrho(p)+p} + f(\Theta) = \hbox{constant},
 \end{equation}
where $f(\Theta)$ is (at this stage) an arbitrary ``function of
integration''.  This is our \emph{preliminary} version of the general
relativistic Bernoulli equation.  To see how we might further simplify
this result, recall (see for example~\cite{LRR, unexpected, ergo})
that for non-relativistic irrotational inviscid barotropic flow the
non-relativistic Bernoulli equation is
 \begin{equation}
 \dot \Theta + {1\over2} (\nabla\Theta)^2 + \int_0^p {\d p\over\rho(p)} + f(t) = \hbox{constant}.
 \end{equation}
(In the non-relativistic case $\rho$ is the mass density.)
Now, in the non-relativistic case we can always redefine
\begin{equation}
\Theta \to \Theta + \int f(t) \d t,
\end{equation}
and use this transformation to eliminate $f(t)$ --- so we can without loss of generality write
 \begin{equation}
 \dot \Theta + {1\over2} (\nabla\Theta)^2 + \int_0^p {\d p\over\varrho(p)} = \hbox{constant}.
 \end{equation}
 In the relativistic case we now note:
 \begin{equation}
-  \log||\nabla \Theta||  + f(\Theta) 
 =
 -  \log(  e^{-f(\Theta)} \; ||\nabla \Theta|| ) 
  =
-    \log|| \nabla F(\Theta)||.
  \end{equation}
That is, in the relativistic case, by making the transformation
 \begin{equation}
 \Theta \to F(\Theta) = \int _0^\Theta e^{-f(\bar\Theta)} \, \d \bar\Theta,
 \end{equation}
we can (\emph{without changing $V^a$}) absorb the arbitrary function
$f(\cdot)$ into a redefinition of $\Theta$, and so write
 \begin{equation}
- \log||\nabla \Theta||
 + \int_0^p {\d p\over\varrho(p)+p}  = \hbox{constant}.
 \end{equation}
Finally, without loss of generality we can rescale $\Theta$ to set the
constant appearing above to zero and so obtain:
 \begin{equation}
\log||\nabla \Theta|| =
\int_0^p {\d p\over\varrho(p)+p}.
 \end{equation}
This is our final form for the general relativistic Bernoulli
equation.  Note that there is no longer any freedom left in choosing
$\Theta$, we have now used it all up.  It is relatively common to
exponentiate the above and rewrite the general relativistic Bernoulli
equation as
 \begin{equation}
||\nabla \Theta|| =
 \exp\left(  \int_0^p {\d p\over\varrho(p)+p} \right).
 \label{E:bernoulli}
 \end{equation}
For the sake of pedagogical development of the argument we have
\emph{not} yet rewritten the integral on the RHS in terms of other
thermodymamic variables; we prefer to delay this for now.

%------------------------------------------------------------------------------------------------------------------------------------------
\subsection{From energy equation to flux conservation}
%------------------------------------------------------------------------------------------------------------------------------------------

The energy conservation equation is
\begin{equation}
\nabla_a ( \varrho V^a) + p (\nabla_a V^a) = 0,
\end{equation}
which becomes
\begin{equation}
V^a \nabla_a\varrho + (\varrho+p) (\nabla_aV^a)=0,
\end{equation}
or
\begin{equation}
V^a \; {\nabla_a\varrho\over\varrho+p}  +  (\nabla_aV^a)=0.
\end{equation}
Using the barotropic condition $\varrho = \varrho(p)$, (but note, now
\emph{without} using the irrotational condition), this can be written
as
\begin{equation}
V^a \nabla_a \left[ \int_{\varrho_{(p=0)}}^\varrho {\d\varrho\over\varrho+p(\varrho)} \right] 
 +  (\nabla_aV^a)=0,
\end{equation}
which implies
\begin{equation}
V^a \nabla_a \left\{ \exp\left[ \int_{\varrho_{(p=0)}}^\varrho {\d\varrho\over\varrho+p(\varrho)} \right] \right\}
 + \exp\left[ \int_{\varrho_{(p=0)}}^\varrho {\d\varrho\over\varrho+p(\varrho)} \right]  (\nabla_aV^a)=0.
\end{equation}
This can now be rewritten as a continuity equation (flux conservation
equation):
\begin{equation}
\nabla_a \left\{ \exp\left[ \int_{\varrho_{(p=0)}}^\varrho {\d\varrho\over\varrho+p(\varrho)} \right]  V^a\right\}=0.
\end{equation}
This is now a ``standard form'' zero-divergence continuity equation
--- note this the ability to derive this transformed form of the
energy equation depends only on the barotropic assumption.  Also note
that we have not yet needed to even introduce, let alone discuss in
any detail, the particle number density $n$, nor introduce any
thermodynamic arguments.

%------------------------------------------------------------------------------------------------------------------------------------------
\subsection{Introducing the number density}
%------------------------------------------------------------------------------------------------------------------------------------------

Subject only to the barotropic condition, we have just derived the
flux conservation equation given immediately above.
We now suppose the barotropic fluid contains some type of conserved
``tracker'' particles. For example, one might be interested in
counting baryons, Fe$^{56}$ nuclei, or leptons.
We now explicitly assume translation invariant and time invariant
\emph{composition} of the fluid. (That is, we assert that the equation
of state is the same throughout the fluid; in other words there is no
explicit time or position dependence in the equation of state.)
Therefore the \emph{ratios} of these tracker particle densities must
be translation and time invariant constants.  Furthermore since these
``tracker particles'' are all assumed to be conserved
\begin{equation}
\nabla_a\left\{ n_i \; V^a \right\} = 0,
\end{equation}
while, since the fluid is assumed to be barotropic, there must be functions of pressure $p$ and energy density $\varrho$ such that
\begin{equation}
n_i = n_i(p) = n_i(\varrho).
\end{equation}
But the only way to satisfy \emph{all} these constraints is if
\begin{equation}
n_i(p) = n_{i\,(p=0)} \; \exp\left[ \int_{\varrho_{(p=0)}}^{\varrho(p)} {\d\varrho\over\varrho+p(\varrho)} \right].
\end{equation}
In particular for the total particle density we have
\begin{equation}
n(p) = n_{(p=0)} \; \exp\left[ \int_{\varrho_{(p=0)}}^{\varrho(p)} {\d\varrho\over\varrho+p(\varrho)} \right].
\end{equation}
This observation is extremely convenient in that allows us to physically interpret the quantity
\begin{equation}
\exp\left[ \int_{\varrho(p=0)}^{\varrho(p)} {\d\varrho\over\varrho+p(\varrho)} \right]   = {n(p)\over n_{(p=0)}} =  {n_i(p)\over n_{i(p=0)}}
\end{equation}
as being proportional to the number density of constituents making up the fluid. (If one wishes to take an extreme point of view and eschew all thermodyamic arguments completely, one could simply take this equation as the definition of a ``shorthand symbol'' $n(p)$, and ignore the physical interpretation of $n(p)$ as particle number density.)
In terms of the number density the conservation equation now simply reads
\begin{equation}
\nabla_a \left\{ n\;  V^a\right\}=0.
\end{equation}
Furthermore, note that using these results we can rewrite the relativistic Bernoulli equation (\ref{E:bernoulli}) as
 \begin{eqnarray}
\log||\nabla \Theta|| &=&
 \int_0^p {\d p\over\varrho(p)+p} =  \int_0^p {\d [\varrho(p) +p]\over\varrho(p)+p} -  \int_{\varrho_(p=0)}^\varrho {\d \varrho\over\varrho+p(\varrho)}
 \\
 &=& \log\left[{\varrho(p)+p\over\varrho_{(p=0)}}\right] - \log\left[{n(p)\over n_{(p=0)}}\right] =  \log\left[{[\varrho(p)+p] \, n_{(p=0)}\over n(p)\,  \varrho_{(p=0)}}\right] .
 \end{eqnarray}
That is
\begin{equation}
||\nabla \Theta|| = {[\varrho(p)+p] \, n_{(p=0)}\over n(p)\, \varrho_{(p=0)}}.
\end{equation}

%------------------------------------------------------------------------------------------------------------------------------------------
\section{Linearization}
%------------------------------------------------------------------------------------------------------------------------------------------
Let us now write
\begin{equation}
\Theta=\Theta_0+\epsilon \,\Theta_1 + \dots,
\end{equation}
which in particular implies that
\begin{equation}
V = V_0 + \epsilon \,V_1 +\dots,
\end{equation}
and further assert
\begin{equation}
\varrho=\varrho_0+\epsilon\, \varrho_1 + \dots,
\end{equation}
\begin{equation}
p=p_0+\epsilon\, p_1 + \dots.
\end{equation}
Using these relations we now  linearize the fluid equations around some assumed background flow. (Note that both the background fluid flow ($V_0$, $\varrho_0$, $p_0$), and the linearized fluctuations, satisfy the Bernoulli and energy conservation equations.)
In a wider context, extending far beyond fluid dynamics,  we mention that it is quite common for linearized fluctuations around an appropriately defined background to exhibit an ``effective spacetime'' behaviour~\cite{LRR, normal1, normal2}.

%------------------------------------------------------------------------------------------------------------------------------------------
\subsection{Linearized continuity equation}
%------------------------------------------------------------------------------------------------------------------------------------------
From the continuity equation we see
\begin{equation}
\nabla_a \left\{ \exp\left[ \int_{\varrho_{(p=0)}}^{\varrho_0+\epsilon\varrho_1+\dots} {\d\varrho\over\varrho+p(\varrho)} \right]  
[V_0^a+ \epsilon V_1^a +\dots]\right\}=0.
\end{equation}
Then to first order in $\epsilon$
\begin{equation}
\nabla_a \left\{ \exp\left[ \int_{\varrho{(p=0)}}^{\varrho_0} {\d\varrho\over\varrho+p(\varrho)} \right]  \;\;
\left(\,{\varrho_1\over\varrho_0+p_0} \; V_0^a+ V_1^a \,\right)\right\}=0.
\end{equation}
Using the number density we can rewrite this as
\begin{equation}
\nabla_a \left\{  n(p_0) \;\;
\left(\,{\varrho_1\over\varrho_0+p_0} \; V_0^a+ V_1^a \,\right)\right\}=0.
\label{E:lin-flux}
\end{equation}

%------------------------------------------------------------------------------------------------------------------------------------------
\subsection{Linearized irrotational flow}
%------------------------------------------------------------------------------------------------------------------------------------------
Perturbing the 4-velocity for relativistic irrotational flow yields
\begin{equation}
V_0 + \epsilon V_1 +\dots ={g^{ab} \; \nabla_b(\Theta_0+\epsilon \Theta_1 + \dots)
\over\sqrt{- g^{ab} \;\nabla_a (\Theta_0+\epsilon \Theta_1 + \dots)\; \nabla_b (\Theta_0+\epsilon \Theta_1 + \dots)}}.
\end{equation}
Then expanding to first order in $\epsilon$ we see
\begin{equation}
V_1^a = {(g^{ab} + V_0^a V_0^b ) \; \nabla_b \Theta_1
\over||\nabla \Theta_0||}.
\end{equation}
Now use the relativistic Bernoulli equation to write
\begin{equation}
V_1^a = (g^{ab} + V_0^a V_0^b ) \;\; \nabla_b \Theta_1\;
\exp\left(  -\int_0^{p_0} {\d p\over\varrho(p)+p} \right).
\end{equation}
Alternatively
\begin{equation}
V_1^a =  {n_0\; \varrho_{(p=0)}\over [\varrho_0+p_0] \; n_{(p=0)}} \;\; (g^{ab} + V_0^a V_0^b ) \;\; \nabla_b \Theta_1.
\label{E:lin-irrotational}
\end{equation}
Note that we automatically have
\begin{equation}
g_{ab} V_1^a V_0^b = 0,
\end{equation}
as required by the normalization condition,
$
g_{ab} V^a V^b = -1,
$
for $V^a$.

%------------------------------------------------------------------------------------------------------------------------------------------
\subsection{Linearized equation of state}
%------------------------------------------------------------------------------------------------------------------------------------------

Linearizing the equation of state we see
\begin{equation}
p_0+\epsilon p_1 +\dots = p(\varrho_0 +\epsilon \varrho_1 +\dots),
\end{equation}
that is
\begin{equation}
p_1 = \left.{\d p\over\d\varrho}\right|_{\varrho_0} \; \varrho_1,
\end{equation}
which we use to define what we shall soon enough see is the speed of sound
\begin{equation}
p_1 = {c_s^2\over c^2} \; \varrho_1.
\end{equation}

%------------------------------------------------------------------------------------------------------------------------------------------
\subsection{Linearized Euler (Bernoulli) equation}
%------------------------------------------------------------------------------------------------------------------------------------------

Linearizing the Bernoulli equation requires us to consider
 \begin{equation}
{1\over2} \log\left[ -g^{cd} \;\nabla_c (\Theta_0+\epsilon \Theta_1 + \dots) \;
\nabla_d(\Theta_0+\epsilon \Theta_1 + \dots) \right]
 = \int_0^{(p_0+\epsilon p_1 + \dots)} {\d p\over\varrho(p)+p}  .
 \end{equation}
Then to first order in $\epsilon$
\begin{equation}
{ -g^{cd} \nabla_c  \Theta_1\nabla_d\Theta_0
\over
 -g^{cd} \nabla_c  \Theta_0\nabla_d\Theta_0} =  {p_1\over\varrho_0+p_0}.
\end{equation}
That is, using the linearized equation of state,
\begin{equation}
-{V_0^a \nabla_a \Theta_1\over||\nabla  \Theta_0||}  = {c_s^2 \over c^2} \; {\varrho_1\over\varrho_0+p_0},
\end{equation}
which we can rearrange to
\begin{equation}
\varrho_1 = - (\varrho_0+p_0){c^2\over c_s^2} {V_0^a \nabla_a \Theta_1\over||\nabla  \Theta_0||}.
\end{equation}
That is, now using the general relativistic Bernoulli equation,
\begin{equation}
\varrho_1 = -(\varrho_0+p_0){c^2 \over c_s^2}  \;
 \exp\left(  -\int_0^{p_0} {\d p\over\varrho(p)+p} \right) 
\; V_0^a \nabla_a \Theta_1.
 \end{equation}
 Finally, using the number density we can rewrite this as
 \begin{equation}
\label{E:lin-bernoulli}
\varrho_1 = -{n_0 \; c^2\over n_{(p=0)} \; c_s^2}  \; \varrho_{(p=0)} \; 
\; V_0^a \; \nabla_a \Theta_1.
 \end{equation}

%------------------------------------------------------------------------------------------------------------------------------------------
\subsection{Deriving the d'Alembertian equation}
%------------------------------------------------------------------------------------------------------------------------------------------

Now combine these results: insert the linearized Euler (Bernoulli)
equation (\ref{E:lin-bernoulli}), and the linearized irrotational
condition (\ref{E:lin-irrotational}), into the linearized continuity
equation (\ref{E:lin-flux}). We obtain
\begin{equation}
\nabla_a \left\{  
    -{n_0^2 \; c^2\; \varrho_{(p=0)} \over n_{(p=0)} \; c_s^2 \; (\varrho_0+p_0)} V_0^a  V_0^b \nabla_b \Theta_1 
   + {n_0^2 \; \varrho_{(p=0)} \over n_{(p=0)} \; (\varrho_0+p_0)}   (g^{ab} + V_0^a V_0^b ) \; \nabla_b \Theta_1
   \right\} = 0.
\end{equation}
But of course $\varrho_{(p=0)}$ and $n_{(p=0)}$ are irrelevant position-independent constants, so we can just as easily write
\begin{equation}
\nabla_a \left\{  
    -{c^2\; n_0^2 \over c_s^2 \; (\varrho_0+p_0)} V_0^a  V_0^b \nabla_b \Theta_1 
   + {n_0^2 \over (\varrho_0+p_0)}   (g^{ab} + V_0^a V_0^b ) \; \nabla_b \Theta_1
   \right\} = 0.
\end{equation}
Introducing the projection tensor $h^{ab} = g^{ab} + V_0^a V_0^b$, and factorizing, this becomes
\begin{equation}
\nabla_a \left\{   {n_0^2\over\varrho_0+p_0} 
\;\;
 \left[   -{c^2\over c_s^2} \; V_0^a  V_0^b  +h^{ab} \right] \; \nabla_b \Theta_1 
   \right\} = 0.
\end{equation}
This is in fact \emph{exactly} the result we want --- a d'Alembertian
equation for the perturbation in the velocity potential $\Theta$.
\begin{itemize}
\item After some trivial notational changes, this agrees  (where the formalisms overlap),  with both the observations of Moncrief~\cite{Moncrief}, and with the presentation of  Bilic~\cite{Bilic}, but the present exposition gives much more attention to the underlying details and makes only a bare minimum of technical assumptions. 
 
\item Note that everything so far is really dimension-independent, and
  that we can now read off the acoustic metric simply by setting
  \def\G{{\mathcal{G}}}
\begin{equation}
\sqrt{-\G}\; \G^{ab} =  {n_0^2\over\varrho_0+p_0} 
\;\;
 \left[   -{c^2\over c_s^2}  \; V_0^a  V_0^b  +h^{ab} \right].
 \end{equation}
The dimension-dependence now comes from solving this equation for
$\G^{ab}$.
\end{itemize}

%------------------------------------------------------------------------------------------------------------------------------------------
\section{The general relativistic acoustic metric}
%------------------------------------------------------------------------------------------------------------------------------------------
\def\G{{\mathcal{G}}}

From the dimension-independent result above
we have, in (d+1) dimensions,
\begin{equation}
(-1)^{(d+1)/2} \; \G^{(d+1)/2-1} =   - {c^2\over c_s^2} \;\;\left( {n_0^2\over\varrho_0+p_0} \right)^{(d+1)},
\end{equation}
whence
\begin{equation}
(-\G)^{(d+1)/2-1} =    {c^2\over c_s^2} \;\;\left( {n_0^2\over\varrho_0+p_0} \right)^{(d+1)},
\end{equation}
Dropping an irrelevant overall constant factor of $c^{-2/(d-1)}$, 
 we finally have the (contravariant) acoustic metric
 \begin{equation}
\G^{ab} = \left( {n_0^2\; c_s^{-1} \over\varrho_0+p_0} \right)^{-2/(d-1)}  \\
 \left\{   -{c^2\over c_s^2}  \; V_0^a  V_0^b  +h^{ab} \right\},
 \end{equation}
 and (covariant) acoustic metric
 \begin{equation}
\G_{ab} =  \;\;\left( {n_0^2\; c_s^{-1}\over\varrho_0+p_0} \right)^{2/(d-1)}  \\
 \left\{  -{ c_s^2\over c^2}  \; [V_0]_a  [V_0]_b  +h_{ab} \right\}.
 \end{equation}
If one wishes to go to the extra trouble of making the acoustic metric dimensionless it is easy to re-insert appropriate position-independent constants in the conformal factor and obtain
 \begin{equation}
\G^{ab} = \left( {n_0^2\; \varrho_{(p=0)}\; c \over n_{(p=0)}^2 \; (\varrho_0+p_0) \; c_s} \right)^{-2/(d-1)}  
 \left\{   -{c^2\over c_s^2}  \; V_0^a  V_0^b  +h^{ab} \right\},
 \end{equation}
 and 
 \begin{equation}
\G_{ab} =   \left( {n_0^2\; \varrho_{(p=0)}\; c \over n_{(p=0)}^2 \; (\varrho_0+p_0) \; c_s} \right)^{2/(d-1)} 
 \left\{  -{ c_s^2\over c^2}  \; [V_0]_a  [V_0]_b  +h_{ab} \right\}.
 \end{equation}
Of course these extra position-independent constants in the conformal factor carry no useful information and are commonly suppressed.

%------------------------------------------------------------------------------------------------------------------------------------------
\section{The non-relativistic limit}
%------------------------------------------------------------------------------------------------------------------------------------------

Compare this general relativistic acoustic metric with the non-relativistic limit, where the coordinates are most conveniently taken to be $x^a = (t;\,x^i)$ and where the d'Alembertian equation reduces to~\cite{LRR, unexpected, ergo}
\begin{equation}
\nabla_a \left\{   {\rho_0} 
\;\;
 \left[   -{1\over c_s^2} \; V_0^a  V_0^b  +h^{ab} \right] \; \nabla_b \Theta_1 
   \right\} = 0,
\end{equation}
with $\rho_0\neq\varrho_0$ now being the \emph{non-relativistic mass
  density}, (not the relativistic energy density), and the meanings of
$V_0$ and $h$ are suitably adjusted. In the non-relativistic case
\begin{equation}
V_0^a = (1; v^i);   \qquad\qquad  
h^{ab} = \left[ \begin{array}{c|c} 0 & 0 \\ \hline 0 & \vphantom{\Big|}\delta^{ij} \end{array} \right],
\end{equation}
and independent of dimensionality we have 
\def\G{{\mathcal{G}}}
\begin{equation}
\sqrt{-\G}\; \G^{ab} =  \rho_0
\;\;
 \left[   -{1\over c_s^2}  V_0^a  V_0^b  +h^{ab} \right],
 \end{equation}
 implying
  \begin{equation}
\G^{ab} = \left( {\rho_0 \over c_s} \right)^{-2/(d-1)}  \\
 \left\{   -{1\over c_s^2}  V_0^a  V_0^b  +h^{ab} \right\},
 \end{equation}
which can be inverted to yield
   \begin{equation}
\G_{ab} = \left( {\rho_0 \over c_s} \right)^{2/(d-1)}  \\
\left[ \begin{array}{c|c} -(c_s^2-v^2) & -v^j \\ \hline -v^i & \vphantom{\Big|}\delta^{ij} \end{array} \right].
 \end{equation}
 
To see how and under what situations the general relativistic acoustic metric reduces to this non-relativistic acoustic metric first consider the conformal factor: In the non-relativistic limit $p_0 \ll \varrho_0$ and $\varrho_0
\approx \bar m\; n_0$, where $\bar m$ is the average mass of the
particles making up to fluid (which by the barotropic assumption is a
time-independent and position-independent constant). So in the
non-relativistic limit we recover the standard result for the
conformal factor~\cite{LRR, Unruh, unexpected, ergo}
\begin{equation}
 {n_0^2\; c_s^{-1}\over\varrho_0+p_0}  \to { n_0\over \bar m c_s} =  {1\over \bar m^2} \;  {\rho_0\over c_s} \propto  {\rho_0\over c_s}.
\end{equation}
 To now probe the tensor structure of the non-relativistic limit, let us recall that:
\begin{itemize}
\item 
$c$ is the speed of light.
\item
$c_s$ is the speed of sound. 
\item
$v$ is the three-velocity of the fluid.
\end{itemize}
Take conventions so that the physical spacetime metric and four-velocity are both
dimensionless. In particular, the coordinates are chosen to be $x^a=(c\,t;\; x^i)$. Now adopting Gaussian normal coordinates at the point of interest
\begin{equation}
g_{ab} = \eta_{ab} = 
\left[\begin{array}{cccc}-1&0&0&0\\0&1&0&0\\0&0&1&0\\0&0&0&1\end{array}\right].
\end{equation}
For the four-velocity we in general have
\begin{equation}
V_0^a = \gamma \left( 1; v^i/c\right); \qquad [V_0]_a = \gamma \left( -1; v^i/c\right).
\end{equation}
Remember that $\d x^0 = c \; \d t$, and note that the $\gamma$ factor is defined using $c$ the physical speed of light. 

Now let us take the nonrelativistic limit.
Ignoring the conformal factor, which comes along for the ride, we have
\begin{equation}
{\cal G}_{ab} \propto \left[ g_{ab} + \left(1-{c_s^2\over c^2}\right) [V_0]_a [V_0]_b \right].
\end{equation}
\begin{itemize}
\item 
Then for the time-time component of the acoustic metric
\begin{equation}
{\cal G}_{00} \propto \left[ -1 + \left(1-{c_s^2\over c^2}\right) \gamma^2 \right].
\end{equation}
This implies
\begin{equation}
{\cal G}_{00} \propto  
\left[ -{c_s^2-v^2\over c^2} + {\mathcal{O}(v^4,c_s^2 v^2)\over c^4} \right].
\end{equation}
So as long as \emph{both} the speed of sound and the 3-velocity of the fluid are small compared to the speed of light we are justified in approximating
\begin{equation}
{\cal G}_{00} \propto -{c_s^2-v^2\over c^2} + \dots.
\end{equation}
\item
In contrast for the time-space components of the acoustic metric 
\begin{equation}
{\cal G}_{0i} \propto \left[ \left(1-{c_s^2\over c^2}\right) \gamma^2 (-1)\left({+v^i\over c}\right)\right].
\end{equation}
This implies
\begin{equation}
{\cal G}_{0i} \propto - \left(1-{\mathcal{O}(c_s^2,v^2)\over c^2}\right)  {v^i\over c}.
\end{equation}
So as long as \emph{both} the speed of sound and the 3-velocity of the fluid are small compared to the speed of light we are justified in approximating
\begin{equation}
{\cal G}_{0i} \propto  - {v^i\over c} + \dots.
\end{equation}
\item
Finally for the space-space components we note
\begin{equation}
{\cal G}_{ij} \propto 
\left[ \delta_{ij} + \left(1-{c_s^2\over c^2}\right) \gamma^2 {v^i \; v^j\over c^2} \right].
\end{equation}
This implies
\begin{equation}
{\cal G}_{ij} \propto 
\left[ \delta_{ij} + \mathcal{O}(v^2/c^2) \right].
\end{equation}
So as long as the 3-velocity of the fluid is small compared to the speed of light we are justified in approximating
\begin{equation}
{\cal G}_{ij} \propto  \delta_{ij} +\dots.
\end{equation}
\end{itemize}
Collecting these results we see that in the nonrelativistic limit
\begin{equation}
{\cal G}_{ab} \propto
\left[\begin{array}{c|c}
-(c_s^2-v^2)/c^2&-v^i/c\\ 
\hline
-v^j/c &\delta_{ij}
\end{array}\right] + \dots
\end{equation}
But now we realise that in the present context  $c$ is just some
convenient fixed conversion constant from $\d x^0= c\,\d t$ to $\d t$, so if we work in terms of the coordinates $(t;\; x^i)$, which are perhaps more natural in the non-relativistic limit, then
\begin{equation}
{\cal G}_{ab} \propto 
\left[\begin{array}{c|c}
-(c_s^2-v^2)&-v^i\\ 
\hline
-v^j&\delta_{ij}
\end{array}\right] +\dots
\end{equation}
as required. Note that to do all this it is essential that \emph{both} $v$ and $c_s$ are
small compared to $c$, though there is no constraint on the relative sizes of $v$ and $c_s$.

%------------------------------------------------------------------------------------------------------------------------------------------
\section{Discussion}
%------------------------------------------------------------------------------------------------------------------------------------------

Under what conditions are the fully general relativistic derivation of
this article necessary? (The non-relativistic analysis of~\cite{LRR,
  Unruh, unexpected, ergo} is after all the basis of the bulk of the
work in ``analogue spacetimes'', and is perfectly adequate for many
purposes.) The current analysis will be needed in four separate
situations:
\begin{itemize}
\item 
When working in an arbitrary curved relativistic
  background;\\ (for example in the problems considered by
  Moncrief~\cite{Moncrief}, and Bilic~\cite{Bilic}).
\item 
Whenever the fluid is flowing at relativistic speeds;\\ (for
  example in the problems considered by Moncrief~\cite{Moncrief}, and
  Bilic~\cite{Bilic}).
\item
Less obviously, whenever the speed of sound is relativistic, even if background  flows are non-relativistic;\\
(for example a near-equilibrum photon gas where $c_s^2 = {1\over3}c^2$ but flow velocities are all small $v \ll c$).
 
\item 
Even less obviously, when the internal degrees of freedom of the
  fluid are relativistic, even if the overall fluid flow and speed of sound are 
  non-relativistic. (That is, in situations where it is necessary to
  distinguish the energy density $\varrho$ from the mass density
  $\rho$; this typically happens in situations where the fluid is
  strongly self coupled --- for example in neutron star cores~\cite{compact} or in
  relativistic BECs~\cite{Stefano-pvt}.)
\end{itemize}
In developing the current derivation, we have tried hard to be clear,
explicit, and \emph{minimal} --- we have introduced only the absolute
minimum of formalism that is requires to do the job, and have eschewed
unnecessary side issues. We hope that the formalism will be useful to
practitioners in the field of ``analogue spacetimes'', particularly
with regard to ongoing and future developments~\cite{Stefano-pvt, Silke-pvt}.  In particular, even in the non-relativistic case it
is already known that adding vorticity greatly complicates the
situation~\cite{vorticity}, and a deeper general relativistic analysis
of this situation would be interesting. Looking further to the future, 
the ``fluid-gravity correspondence'' hints at even deeper connections between curved spacetimes and fluid dynamics~\cite{fluid-gravity, fg2, fg3}. 

%------------------------------------------------------------------------------------------------------------------------------------------
\section*{Acknowledgements}
%------------------------------------------------------------------------------------------------------------------------------------------

This research was supported by the Marsden Fund administered by the
Royal Society of New Zealand. The authors would like to thank Silke
Weinfurtner, Bill Unruh, Stefano Liberati, Serena Fagnocchi, and
Stefano Finazzi for their comments and suggestions.

%------------------------------------------------------------------------------------------------------------------------------------------
\appendix
%------------------------------------------------------------------------------------------------------------------------------------------
\section*{Appendix: Specific enthalpy and thermodynamic considerations}
%------------------------------------------------------------------------------------------------------------------------------------------

We now consider some thermodynamics which for pedagogical purposes we
have delayed as much as possible.  (Though delayed, we emphasize that thermodynamics does have important consequences --- see for instance~\cite{Grisha}.) Suppose we take the specific enthalpy as primary
\begin{equation}
w = {\varrho+p\over n},
\end{equation}
and use the fact that we have already deduced
\begin{equation}
n = n_{(p=0)} \; \exp\left[ \int_{\varrho_{(p=0)}}^\varrho {\d\varrho\over\varrho+p(\varrho)} \right],
\end{equation}
to then write
\begin{equation}
w = \exp[\log(\varrho+p)] n^{-1} = \exp\left\{ \int {\d [\varrho+p]\over \varrho+p}\right\} 
 \; \exp\left[ - \int_{\varrho_{(p=0)}}^\varrho {\d\varrho\over\varrho+p(\varrho)} \right].
\end{equation}
This implies
\begin{equation}
w =  {\varrho_{(p=0)}\over n_{(p=0)}} \exp\left[  \int_{0}^p {\d p\over \varrho+p} \right] = w_{(p=0)}\; \exp\left[  \int_{0}^p {\d p\over \varrho+p} \right].
\end{equation}
This so far is a purely (barotropic) thermodynamic result, independent
of any irrotational condition. 

A secondary result, now specifically tied to the Bernoulli equation
(and hence to irrotational flow), is that
\begin{equation}
w = w_{(p=0)} \; ||\nabla\Theta||.
\end{equation}
Note that the way we have presented the derivation we have been able
to delay and avoid the need for thermodynamic arguments as far as
possible.  (In contrast, this equation is the starting point adopted
by Bilic in his derivation of relativistic acoustic
geometry~\cite{Bilic}, what for us is a peripheral result has in that
analysis moved to centre stage --- and gives we feel far too central a
role to thermodynamic issues.)

In a similar vein, the energy equation
\begin{equation}
\nabla_a ( \varrho V^a) + p (\nabla_a V^a) = 0,
\end{equation}
can be combined with the conservation equation
\begin{equation}
\nabla_a ( n V^a) = 0,
\end{equation}
as follows:
\begin{equation}
\nabla_a V^a = - {1\over \varrho+p} {V^a \nabla_a \varrho} = - {1\over \varrho+p } \;{\d \varrho\over\d\tau},
\end{equation}
\begin{equation}
\nabla_a V^a = - {1\over n} {V^a \nabla_a n} = - {1\over n} \; {\d n\over \d\tau},
\end{equation}
where $\d/\d\tau$ now refers to a material derivative along the
flow. Eliminating the divergence
we have
\begin{equation}
 {1\over \varrho+p } \; {\d \varrho\over\d\tau} = {1\over n} \; {\d n\over \d\tau}.
 \end{equation}
More formally, by invoking the barotropic condition we see that for
any ``fluid element'' (\emph{i.e.}, tiny lump of fluid) we have
\begin{equation}
 {\d\varrho \over \varrho+p }  = {\d n\over n},
\end{equation}
which can be rearranged as
\begin{equation}
\d (\varrho/n) = - p \; \d(1/n).
\end{equation}
\def\V{{\mathcal{V}}} 
In terms of $\V$, the ``specific volume" of a
fluid element, we have $\V\propto 1/n$ and so
\begin{equation}
\d (\varrho\;\V) = - p \; \d\V,
\end{equation}
which connects back to basic thermodynamics and again clearly verifies
that in a relativistic setting $\varrho$ is the energy density, while
$n$ is the number density.

\bigskip
\noindent
Finally, in view of these comments we can now write
\begin{equation}
\sqrt{-\G}\; \G^{ab} =  {n_0\over w_0} 
\;\;
 \left[   -{c^2\over c_s^2}  \; V_0^a  V_0^b  +h^{ab} \right],
 \end{equation}
 and somewhat simplify the acoustic metric to read
  \begin{equation}
\G^{ab} = \left( {n_0\over w_0 \;  c_s} \right)^{-2/(d-1)}  \\
 \left\{   -{c^2\over c_s^2} \; V_0^a \, V_0^b  +h^{ab} \right\};
 \end{equation}
 \begin{equation}
\G_{ab} =  \;\;\left( {n_0\over w_0  \; c_s} \right)^{2/(d-1)}  \\
 \left\{   -{ c_s^2\over c^2}  \; [V_0]_a  \, [V_0]_b  +h_{ab} \right\}.
 \end{equation}

%------------------------------------------------------------------------------------------------------------------------------------------

%------------------------------------------------------------------------------------------------------------------------------------------
%------------------------------------------------------------------------------------------------------------------------------------------
\end{document}